\pdfoutput = 1
\documentclass[apj,numberedappendix]{emulateapj}

\usepackage{epsfig}
\usepackage{amsmath}
\usepackage{enumerate}
\usepackage{mathrsfs}
\usepackage{natbib}
\usepackage{units}
\usepackage{hyperref}

\usepackage{ulem}
\usepackage[squaren, Gray, cdot]{SIunits}

\usepackage{color}
\definecolor{orange}{RGB}{220,110,0}

\newcommand{\md}{\mathrm{d}}
\newcommand{\alfven}{Alfv\'{e}n }

\newcommand{\bb}{\boldsymbol{B}}
\newcommand{\be}{\boldsymbol{E}}
\newcommand{\bj}{\boldsymbol{j}}

\newcommand{\bB}{\boldsymbol{B}}
\newcommand{\bE}{\boldsymbol{E}}
\newcommand{\vd}{v_{\rm d}}
\newcommand{\vdx}{v_{{\rm d},x}}
\newcommand{\Sabs}{S_{\rm abs}}

\begin{document}

\title{
Fast dissipation of Colliding Alfv\'en Waves in a Magnetically Dominated Plasma}

\author{Xinyu Li\altaffilmark{1,2}}
\author{Andrei M. Beloborodov\altaffilmark{3,4}}
\author{Lorenzo Sironi\altaffilmark{5}}

\affil{$^1$Canadian Institute for Theoretical Astrophysics, 60 St George St, Toronto, ON M5R 2M8\\
$^2$Perimeter Institute for Theoretical Physics, 31 Caroline Street North, Waterloo, Ontario, Canada, N2L 2Y5 \\
$^3$Department of Physics and Columbia Astrophysics Laboratory, Columbia University, 
538 West 120th Street, New York, NY 10027\\
$^4$Max Planck Institute for Astrophysics, Karl-Schwarzschild-Str. 1, D-85741, Garching, Germany \\
$^5$Department of Astronomy and Columbia Astrophysics Laboratory, Columbia University, 
538 West 120th Street, New York, NY 10027\\}

\begin{abstract}
\medskip
Magnetic energy around  compact objects often dominates over plasma rest mass, and its dissipation can power the object luminosity. 
We describe a dissipation mechanism which works faster than magnetic reconnection. 
The mechanism involves two strong Alfv\'en waves with anti-aligned magnetic fields $\bB_1$ and $\bB_2$ that propagate in opposite directions along background magnetic field $\bB_0$ and collide.
The collision forms a thin current sheet perpendicular to $\bB_0$, which absorbs the incoming waves.
The current sheet is sustained by electric field $\bE$ breaking the magnetohydrodynamic condition $E<B$ and accelerating particles to high energies.
We demonstrate this mechanism with kinetic plasma simulations using a simple setup of two symmetric plane waves with amplitude $A=B_1/B_0=B_2/B_0$ propagating in a uniform $\bB_0$.
The mechanism is activated when $A>1/2$.
It dissipates a large fraction of the wave energy, $f=(2A-1)/A^2$, reaching 100\% when $A=1$. 
The plane geometry allows one to see the dissipation process in a one-dimensional simulation. 
We also perform two-dimensional simulations, enabling spontaneous breaking of the plane symmetry by the tearing instability of the current sheet. 
At moderate $A$ of main interest the tearing instability is suppressed.
Dissipation  transitions to normal, slower, magnetic reconnection at $A\gg 1$.
The fast dissipation described in this paper may occur in various objects with perturbed magnetic fields, including magnetars,  jets from accreting black holes, and pulsar wind nebulae.
\end{abstract}


\section{Introduction}
Fast dissipation of magnetic energy is a key process feeding the luminosity of strongly magnetized objects. 
A canonical example is magnetars, which produce short powerful X-ray bursts and giant flares (see \citet{2017ARA&A..55..261K} for a review). 
Two possible  mechanisms of fast dissipation in magnetars have been studied: a magnetospheric turbulence cascade \citep{1998PhRvD..57.3219T, 2019ApJ...881...13L} and magnetic reconnection \citep{1996ApJ...473..322T,2003MNRAS.346..540L,2013ApJ...774...92P,2020ApJ...900L..21Y,2020arXiv201107310B}. 
It is not established which mechanism dominates the observed activity. 

A similar open question concerns the origin of fast dissipation in other systems, including magnetically dominated coronae and jets of accreting black holes and the Crab nebula.
Particularly challenging are the ultra-fast gamma-ray flares occasionally observed in these systems, indicating sudden dissipation events, perhaps associated with magnetohydrodynamic instabilities. 
It was also proposed that fast dissipation can occur in specially prepared unstable magnetic configurations, with immediate onset of violent magnetic reconnection \citep{2016ApJ...826..115N,2018JPlPh..84b6301L}. 

In this paper, we describe a different dissipation mechanism which naturally occurs in magnetically dominated systems with strong waves. 
It works quickly, on the light crossing timescale, and its efficiency can reach $100\%$. 
Alfv\'en waves in a magnetically dominated plasma propagate with nearly speed of light, and the wave electric field $E$ can approach the magnetic field $B$, in particular when two waves collide. When $E$ reaches $B$, the ideal magnetohydrodynamics (MHD) breaks, which causes immediate strong dissipation.
The conditions for triggering this mechanism can be seen analytically by considering two symmetric counter-propagating Alfv\'en wave packets with linear polarizations. Propagation of an isolated packet may be well described in the framework of ideal force-free electrodynamics (FFE), which neglects plasma inertia. 
However, the packet collision becomes a kinetic plasma problem, because the collision is dissipative and we wish to see how the particles gain energy.

In Section~\ref{sec:theory}, we give an analytical description of the wave collision, using a simple setup of two symmetric plane waves propagating along the $x$ axis. 
Then, in Section~\ref{sec:simulation}, we employ direct kinetic simulations to demonstrate the dissipation mechanism.
We first perform one-dimensional (1D) simulations, with symmetry in the $y$ and $z$ directions, and then simulate the same system in two dimensions (2D) allowing the development of instabilities along $y$ (the direction of the wave magnetic field). 
Our results are summarized in Section~\ref{sec:discussion}.

\section{Analytical Description}\label{sec:theory}
\subsection{Setup} \label{setup}
The simplest setup of the wave collision problem is shown in the left panel of Figure~\ref{fig:squarewave}. 
Two wave packets with a square profile are propagating in the $\pm x$ directions and collide at $x=0$. 
All quantities are independent of $y$ and $z$, i.e. we deal with plane waves where the electromagnetic fields $\bE$ and $\bB$ are functions of $x$ and time $t$. 
The constant background magnetic field $\bB_0$ is uniform and directed along the $x$ axis.
The two waves are linearly polarized. 
The best case for strong dissipation is where the two waves have equal electric fields $\bE=(0,0,E_z)$.
The corresponding magnetic field is $\bB=(B_0,B_y,0)$, with $B_y=\pm E_z$ having opposite signs for the two waves. 

We will assume that the background plasma is strongly magnetically dominated, i.e. it has the magnetization parameter
\begin{equation}
    \sigma_0\equiv\frac{B_0^2}{4\pi n_0 mc^2}\gg 1,
\end{equation}
where $n_0$ is the number density of plasma particles carried by the waves, and $m$ is the particle mass. 
For simplicity, we will assume that the $\pm$ charges have the same mass $m$. 
This is satisfied for electron-positron plasma, which is expected in the astrophysical objects of interest. 

In the limit of $\sigma_0\rightarrow\infty$, the wave propagates with the speed of light $c$, without any distortion.
As long as the two wave packets remain isolated, i.e. before the collision, their propagation does not involve excitation of electric current: the wave electric field $\bE\perp\bB_0$ is unable to move the magnetized $e^\pm$ particles and excite a current. 
The propagation occurs as if the magnetized plasma was replaced by vacuum.
Such waves can be viewed as \alfven waves with wavevectors parallel to the background field. 
In this special case, the \alfven wave becomes degenerate with the other FFE mode, the fast magnetosonic wave.

The dimensionless wave amplitude is defined by
\begin{equation}
    A\equiv\frac{|E_z|}{B_0}=\frac{|B_y|}{B_0}.
\end{equation}
The two wave packets will be assumed to have the same amplitude.
There is no loss of generality in this assumption, because for incoming waves with different amplitudes one can always choose a new inertial frame (boosted along the $x$ direction) where the amplitudes become equal.

The two initial packets have the Poynting fluxes
\begin{equation}
    \boldsymbol{S}=c\,\frac{\bE\times\bB}{4\pi}=\pm c\, \frac{E^2}{4\pi}\,\boldsymbol{e}_x-c\,\frac{EB_0}{4\pi}\,\boldsymbol{e}_y, 
\end{equation} 
where $(\boldsymbol{e}_x,\boldsymbol{e}_y,\boldsymbol{e}_z)$ are the unit vectors of the Cartesian coordinate system. 
We will denote the magnitude of the initial Poynting flux toward the collision center $x=0$ by $S_0=|S_x|$,
\begin{equation}
  S_0=\frac{c}{4\pi}\,A^2 B_0^2.
\end{equation}
The total (free) wave energy per unit area is defined by
\begin{equation}
    U = \int \frac{E^2+B^2-B_0^2}{8\pi}\;\md x.
\end{equation}
It excludes the contribution of the constant background magnetic field.

It is convenient to define two characteristic frequencies of the problem,
\begin{equation}
    \omega_p\equiv \sqrt{\frac{4\pi e^2 n_0}{m}}, \qquad \omega_B\equiv \frac{eB_0}{mc}=\sqrt{\sigma_0}\,\omega_p.
\end{equation}
We will assume that the length of the wave packet $\lambda$ is much larger than the characteristic plasma scale $c/\omega_p$. 
As will be shown below, $c/\omega_p$ sets the characteristic width $\Delta$ of the dissipation layer (current sheet) established during the wave collision.

\subsection{Wave Dynamics}
The dynamics of electromagnetic field is governed by the Maxwell's equations,
\begin{eqnarray}
\frac{\partial \bb}{\partial t} &=& -c\nabla\times\be, \label{eqn:induction}\\
\frac{\partial \be}{\partial t} &=& c\nabla\times\bb -4\pi \bj.\label{eqn:ampere}
\end{eqnarray}
The vacuum-like propagation occurs as long as $\bj=0$.

\begin{figure*}
\includegraphics[width=0.98\textwidth]{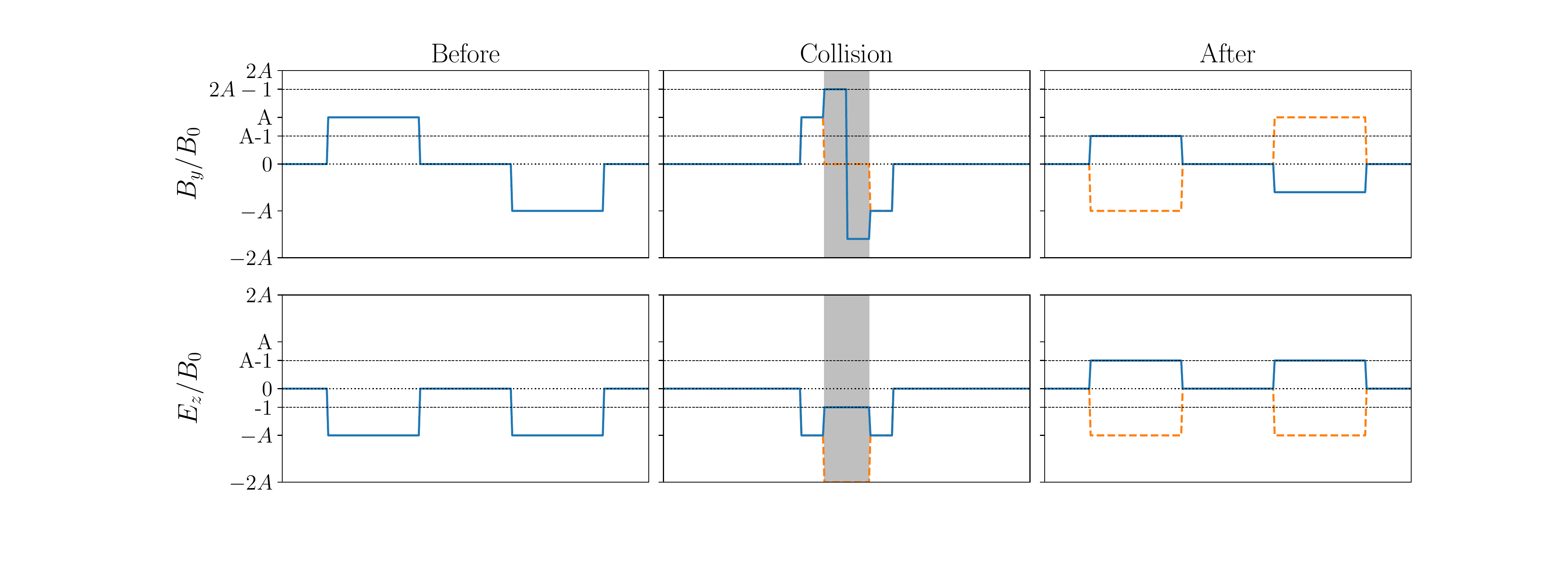}
\caption{
Collision of two symmetric plane  waves with equal amplitudes $A$ ($A>1$ is assumed in the figure).
Left: the fields in the two wave packets before the collision. 
Middle: the fields during the collision (blue). 
Two reflected waves with $E_z=B_0$ become superimposed on the incoming waves in a growing region (shaded in grey). 
The jump of $B_y$ at the center has changed from $\pm AB_0$ to $\pm(2A-1)B_0$. 
For comparison, the dashed orange line shows how the two packets would propagate through each other in vacuum (this would apply if $A<1/2$).
Right: the two packets after the collision (blue). 
The reflection with amplitude $A-1$ is now complete and the reflected packets are moving away from the center. 
The final state is different from free propagation of the packets through each other (orange).
}
\label{fig:squarewave}
\end{figure*}

Consider first what would happen if the current density $\bj$ remained everywhere zero when the two wave packets approach and overlap.
Then, the waves would continue their linear evolution and keep propagating with speed $c$. 
The electromagnetic field would be described simply by the superposition of the two waves, as shown by the orange dashed lines in Figure~\ref{fig:squarewave}.
Where the two waves overlap, their opposite magnetic fields cancel, and their equal electric fields add up, so 
\begin{equation}
    B_y=0,\qquad |E_z|=2AB_0. 
\end{equation}
The overlapping region expands with speed $2c$ until the two symmetric packets totally overlap. 
Then, the overlapping region begins to shrink and eventually vanishes, after the two waves pass through each other and exit the collision with no change from their original form.
In the ovelapping region we find
\begin{equation}
    E^2-B^2 = B_0^2(4A^2-1) \qquad
    \left(\textrm{for} \; A<\frac{1}{2}\right).
\end{equation}
This relation holds as long as $E<B$, which requires $A<1/2$. 

For amplitudes $A>1/2$, the magnetic dominance is broken. 
Then, the collision cannot occur with $\bj=0$. 
Plasma particles exposed to $E>B$ are no longer magnetized, and $e^\pm$ are accelerated by $\bE$ along $\pm z$, creating an electric current $\bj=(0,0,j_z)$. 
The current excitation will buffer the growth of $\bE$ in the collision region. 
It tends to screen $\bE$ similar to the well known screening of low frequency electromagnetic waves in unmagnetized plasma. 
The screening occurs on the short timescale $\sim\omega_p^{-1}\ll \lambda/c$, i.e. immediately after reaching $E>B$. 
It prevents any significant growth of $E>B$, so that $E$ in the collision region is bound by
\begin{equation}
  E_{\rm screen}=B \qquad \left(\textrm{for} \; A>\frac{1}{2}\right).
\end{equation}
Note that $B_y=0$ at $x=0$ at all times, by symmetry. 
Therefore, $B=B_0$ at the collision center. 
Thus, the presence of plasma imposes a simple boundary condition for the wave collision problem,
\begin{equation}
    |E_z|=B_0,\qquad B_y=0 \qquad (x=0).
\end{equation}

This condition allows one to quickly find the evolution of the electromagnetic field during the packet collision. 
The plane $x=0$ reflects the incoming waves with amplitude $\bE_r$ such that the total $E=B_0$. 
It turns out sufficient for the system to create a thin current sheet at $x=0$ (where it is easiest to break magnetic dominance), and outside the current sheet the reflected waves propagate in the linear, vacuum regime. 
The validity of magnetic dominance $B>E$ (and hence $j=0$) can be verified after the wave reflection solution is obtained from the boundary value problem. 
This solution will also be verified below by the direct kinetic simulation of the plasma behavior in the colliding waves.  

The problem is symmetric about the plane $x=0$, so two symmetric reflected waves will be launched, with equal amplitudes $A_r=|\bE_r|/B_0$. 
They propagate away from $x=0$ on top of the two original incoming waves with amplitude $A$. 
The reflected wave has electric field $\bE_r=(0,0,E_{z,r})$ such that the total electric field satisfies $E=B_0$. This gives
\begin{equation}
  A_r=|A-1|.
\end{equation}
$E_{z,r}$ has the same sign as the incoming wave if $A<1$, and changes sign if $A>1$. 
The magnetic field of the reflected wave $\bB_r$ satisfies $|B_{y,r}|=|E_{z,r}|$, and its sign is such that the wave propagates away from $x=0$: $\bE_r\times\bB_r$ is along $\boldsymbol{e}_x$ at $x>0$ and along $-\boldsymbol{e}_x$ at $x<0$. 

Note that the reflected wave vanishes when $A=1$; in this special case the incoming waves are completely absorbed at the collision centre, yielding the maximum dissipation efficiency. 
This will be confirmed below by the 1D and 2D kinetic simulations. The limit of $A\gg 1$ gives $A_r/A\approx 1$, suggesting an ``elastic'' collision -- the two packets bounce from each other with negligible absorption. 
This elastic behavior holds only in the 1D model. 
We will show in Section~\ref{sec:sim2d} that the 1D description fails when $A\gg 1$; then a different dissipation mode is activated by the tearing instability of the current sheet.

The superposition of the incoming and reflected waves determines the electromagnetic field during the collision of packets with $A>1/2$.
It is shown in Figure~\ref{fig:squarewave} for the case of $A>1$.
After the collision is over, only the reflected waves are left.
Note that $A>1$ gives the fields in the final state with signs opposite to the vacuum solution that was found for $A<1/2$. 
When the initial amplitude satisfies $1/2<A<1$, the final fields have the same signs as the original waves, as if the packets have propagated through each other and came out with a reduced amplitude $|A-1|=1-A$. 
When $A\rightarrow 1/2$ the solution obtained using the reflected waves becomes identical to free propagation of the initial waves through each other.

During the collision of packets with $A>1/2$, $B_y$ in the interaction region (the grey stripe in Figure~1) does not cancel to zero. 
The superposition of the incoming and reflected waves gives
\begin{equation}
\label{eq:By}
    |B_y|=(2A-1)B_0,
\end{equation}
and $B_y$ has the opposite signs at $x>0$ and $x<0$.
Equation~(\ref{eq:By}) implies that $B>E$ everywhere except the narrow layer at $x=0$ where $B_y$ changes sign. 
The excitation of electric current becomes possible in this layer because $B_y=0$ (at $x=0$) allows $E$ to approach $B$. 
The surface electric current sustaining the jump of
$B_y$ at $x=0$ is given by
\begin{equation}
\label{eq:I}
I=\frac{c|B_y|}{2\pi} = \frac{(2A-1)cB_0}{2\pi}.
\end{equation}
Outside the narrow central layer the waves propagate as in vacuum ($j=0$), giving the simple linear superposition of the incoming and reflected waves.

The fact that the reflected wave amplitude is reduced from $A$ to $|A-1|$ implies that the following fraction $f$ of the incoming Poynting flux $S_0$ is absorbed at $x\approx 0$, 
\begin{equation}\label{eqn:frac_dis}
    f=\frac{A^2-(A-1)^2}{A^2}=\frac{2A-1}{A^2}.
\end{equation}
The power absorbed per unit area, from each side $\pm x$, is $fS_0=fA^2B_0^2c/4\pi$. 
The electromagnetic power absorbed from both sides equals the power dissipated in the current sheet,
\begin{equation}
\label{eq:power}
    2fS_0=EI=B_0I.
\end{equation}
This self-consistency check completes the basic description of the electromagnetic field during the wave collision. 

\subsection{Particle Dynamics}
\subsubsection{Particle Motion outside the Current Sheet}
Let us first consider particle motion in the isolated wave, before the collision begins.
The wave packet has uniform orthogonal fields $\bE$ and $\bB$, and the particle motion in this situation is well known (Appendix~\ref{sec:app_motion1}).
The electromagnetic field of the packet satisfies 
\begin{equation}
\frac{E}{B}=\frac{|E_z|}{(B_0^2+B_y^2)^{1/2}}=\frac{A}{\sqrt{1+A^2}}<1.
\end{equation}
Therefore, the particle motion on timescales $\gg\omega_{B}^{-1}$ is a drift in the direction of the Poynting flux with velocity   $\boldsymbol{v}_{\rm d}=c\,\bE\times\bB/B^2$. 
Note that $\bE\times\bB=(-E_zB_y,E_zB_0,0)$ has the non-zero $x$ and $y$ components. 
The Poynting flux and the particle drift in the $y$ direction are not causing any interesting transport, because the system is symmetric under translation along $y$. 
The drift speeds along $x$ in the isolated packets before the collision are
\begin{equation}
    \vdx^0 = \frac{E_zB_y}{B^2}c=\pm\frac{A^2}{A^2+1}\,c.
\end{equation}
The particles are drifting in the direction of the wave packet propagation toward the future collision center.
The Lorentz factor of the particles is
\begin{equation}
    \gamma_{\rm d}^0 = \frac{1}{\sqrt{1-(\vd^0/c)^2}}  
    = \sqrt{A^2+1}.
\end{equation}

We consider here waves in a magnetically dominated plasma, so that the group Lorentz factor of the wave, $\gamma_{\rm gr}\approx \sigma_0^{1/2}$ can be taken as infinity in the first approximation, i.e. the wave packet propagates with speed $c$.
The particle drifts relative to the packet with speed $|\vdx^0|-c=-c/(A^2+1)$, and it takes time 
\begin{equation}
  t_{\rm cross}=\frac{\lambda}{c-|\vdx^0|}=(A^2+1)\,\frac{\lambda}{c}
\end{equation} 
for the particle to cross the packet of length $\lambda$ and exit behind it. 

The particle number density inside the packet, $n_0$, obeys the continuity equation. 
For instance, consider the packet propagating in the $+x$ direction. Using the coordinate $\xi=x-ct$, the continuity equation for the plasma with density $n(\xi,t)$ and speed $v_x(\xi,t)$ reads 
\begin{equation}
    \frac{\partial n}{\partial t}+\frac{\partial}{\partial \xi}\left[n(v_x-c)\right]=0.
\end{equation}
In a steady state ($\partial/\partial t=0$), this yields 
\begin{equation}
\label{eq:n}
    n_0 = \frac{n_b}{1-\vdx^0/c}=(A^2+1)n_b,
\end{equation}
where $n_b$ is the unperturbed plasma density ahead (and behind) the wave packet, where  $v_x=0$.

After the collision begins, the particles swept by the packets with $A>1/2$ will encounter the reflected wave. 
Then they become exposed to the electromagnetic field  that satisfies $|E_z|=B_0$ and $|B_y|=(2A-1)B_0$, and continue to drift toward the center with a reduced speed
\begin{equation}
\label{eq:vdx}
\vdx=\pm\frac{|E_z B_y|}{B_0^2+B_y^2}c
=\frac{(2A-1)}{4A^2-4A+2}c.
\end{equation}
The density of the particles coming toward $x=0$ is then changed to 
\begin{equation}
  n=\frac{c+|\vdx^0|}{c+|\vdx|}\,n_0.
\end{equation}
Note that the jump of the drift speed from $\vdx^0$ to $\vdx$ across the front of the reflected wave can be thought of as a shock wave. 
The shock is mildly relativistic and the energy it can dissipate per particle is small compared to the power released in the current sheet at the collision center (Equation~\ref{eq:power}), which scales with $\sigma_0\gg 1$.

\subsubsection{Particle Motion inside the Current Sheet}
We now examine in more detail the plasma behavior in the dissipative collision of waves with $A>1/2$.
The particle number flux from each side toward $x=0$ is
\begin{equation}
    F=n\,|\vdx|=\frac{(2A^2+1)(2A-1)}{(A^2+1)(4A^2-2A+1)}\,cn_0,
\end{equation}
and the Poynting flux absorbed at the collision center is
\begin{equation}
    \Sabs = fS_0= \frac{c}{4\pi} (2A-1)B_0^2.
\end{equation}
This implies that on average the particles passing through the dissipation region gain the Lorenz factor
\begin{equation}
\label{eq:gam_av}
    \bar{\gamma}\approx\frac{\Sabs}{Fmc^2}\approx \frac{(4A^2-2A+1)(A^2+1)}{2A^2+1}\,\sigma_0,
\end{equation}
where we have neglected the particle Lorentz factor $\gamma\sim 1$ before the collision.

The conversion of Poynting flux $\Sabs$ to plasma energy occurs in the thin layer $|x|<\Delta$ occupied by the current sheet, where $B_y$ changes from $bB_0$ to $-bB_0$ ($b= 2A-1$, see Equation~(\ref{eq:By})). 
The motion of particles entering the dissipation layer is described in Appendix~\ref{sec:app_motion}. 
The profile of the magnetic jump $B_y(x)$ and its half-width $\Delta$ self-organize so that the particle motion in the layer self-consistently sustains $j_z=(c/4\pi)dB_y/dx$. 
As discussed in Appendix~\ref{sec:app_motion}, this is achieved when
\begin{equation}
\label{eq:Delta0}
    \frac{b\omega_B\Delta}{\bar{\gamma}^{1/2} c}\sim 1.
\end{equation}
Then, using Equation~(\ref{eq:gam_av}), we find
\begin{equation}
\label{eq:Delta}
    \Delta\sim \frac{c}{\omega_p}.
\end{equation}

The time spent by the particles in the dissipation layer is (see Equation~(\ref{eq:delta_t1}))
\begin{equation}
\label{eq:delta_t}
    \delta t\sim \frac{ \bar{\gamma}^{3/2}}{\omega_B},
\end{equation}
and the column plasma density in the layer may be estimated as 
\begin{equation}
    N\sim 2F\,\delta t\sim 2\,\frac{\Sabs \gamma^{1/2}}{mc^2\omega_B} \sim \frac{(2A-1)B_0}{2\pi e}\,\bar{\gamma}^{1/2}.
\end{equation}
The net electric current in the layer is $I\sim eN|\bar{v}_z|$, where $\bar{v}_z\sim - c/\bar{\gamma}^{1/2}$ (Equation~\ref{eq:vz_av}). 
This gives $I$ consistent with Equation~(\ref{eq:I}), as expected.

The steady-state density of accelerated particles inside the dissipation layer is 
\begin{equation}
\label{eq:nacc}
    n_{\rm acc}\sim \frac{N}{\Delta}\sim \bar{\gamma}\, n_0.
\end{equation}
The accelerated particles continually spread from the layer in the $\pm x$ directions with velocity $v_x\sim\pm c/b\bar{\gamma}$ (Equation~\ref{eq:vx}), and by the end of the packet collision occupy the region $|x|\lesssim \lambda/\bar{\gamma}\ll \lambda$.
This dense slab of accelerated particles contains the energy dissipated by the wave collision.
\footnote{After the collision ends, the current $I$ decays and emits additional waves, returning some energy of the accelerated particles to the electromagnetic field.
When $\lambda\gg \bar{\gamma}\Delta$, this residual emission phase is short compared to the main, quasi-steady phase of the packet collision, and it weakly affects the net dissipation efficiency.}
We are particularly interested in  macroscopic wave packets with
\begin{equation}
    \lambda\gg \bar{\gamma}\,\Delta.
\end{equation} 
Then the collision duration$\lambda/c\gg \delta t$, so dissipation occurs quasi-steadily, and the slab occupied by the energized particles is much thicker than the dissipation layer $\Delta$.


\section{Kinetic Simulations} \label{sec:simulation}
We perform kinetic simulations of wave collision with the particle-in-cell code TRISTAN-MP \citep{2005AIPC..801..345S}.
The code calculates the evolution of $\bE$ and $\bB$ fields from Maxwell equations discretized on a Cartesian grid. 
The electric current in each grid cell is found by following the motions of individual particles.

The setup of initial conditions is similar for the 1D and 2D simulations.
It represents two counter-propagating plane waves, as described in Section~\ref{setup} and illustrated in the left panel of Figure~\ref{fig:squarewave}.
Our fiducial model has the magnetization parameter $\sigma_0=20$.
The box is initially filled with cold neutral $e^\pm$ plasma moving with the drift velocity $c\,\bE\times\bB/B^2$.
The plasma has a uniform density $n_0=n_++n_-$, which defines the reference plasma frequency $\omega_p=(4\pi e^2 n_0/m)^{1/2}$.
The initial density $n_0$ is represented by $64$ particles per cell.

In all 1D simulations (and most of 2D simulations) the wave packets have the size of $\lambda=3072\,c/\omega_p$. The square profile of each packet is smoothed at the edges by adding exponential wings of width $ 50\,c/\omega_p$. 
The computational box has $51200$ cells in the $x$ direction, with $8$ cells per skin depth $c/\omega_p$.
The box size is chosen slightly larger than $2\lambda$, so that it accommodates the two packets.
The choice of boundary conditions at the $x$ boundaries of the box is not important, because the packet collision ends before the waves emerging from the collision reach the boundaries.
The plasma near the boundaries remains at rest throughout the simulation, and we use the simple periodic boundary conditions. 

The initial separation between the packets (not including the wings) is $128\,c/\omega_p$.
The main collision phase begins when the flat parts of the two packets approach $x=0$, which occurs after time $\sim 64\omega_p^{-1}$. We choose $t_0=-64\,\omega_p^{-1}=-0.02\,\lambda/c$ at the start of the simulation, so that $t=0$ approximately corresponds to the beginning of the strong packet collision.

\subsection{1D Simulations} \label{sec:sim}
In the 1D simulations, the electromagnetic field is not allowed to spontaneously develop any variations in the $y$ or $z$ directions during the wave collision.
In addition, the symmetry of $e^\pm$ motions in the symmetric setup of the wave collision implies $E_x=B_z=0$ throughout the 1D evolution. 
Numerical noise caused by the finite number of particles can violate this condition, and then fluctuations of $E_x$ and $B_z$ can grow over time.  
These fluctuations should disappear in the limit of $n_0 c/\omega_p\rightarrow \infty$, which is adequate for most astrophysical objects, however this fluctuation-free limit is expensive to simulate. 
Instead, we simply impose $E_x=B_z=0$ in the 1D simulations.\footnote{We also performed simulations where numerical fluctuations of $E_x$ and $B_z$ are allowed to grow.
These noisier numerical models demonstrate similar dissipation and particle acceleration during the wave collision. 
We verified that the spontaneous growth of $E_x$ and $B_z$ is reduced with increasing number of particles per cell.
}

\begin{figure*}[t]
\includegraphics[width=\textwidth]{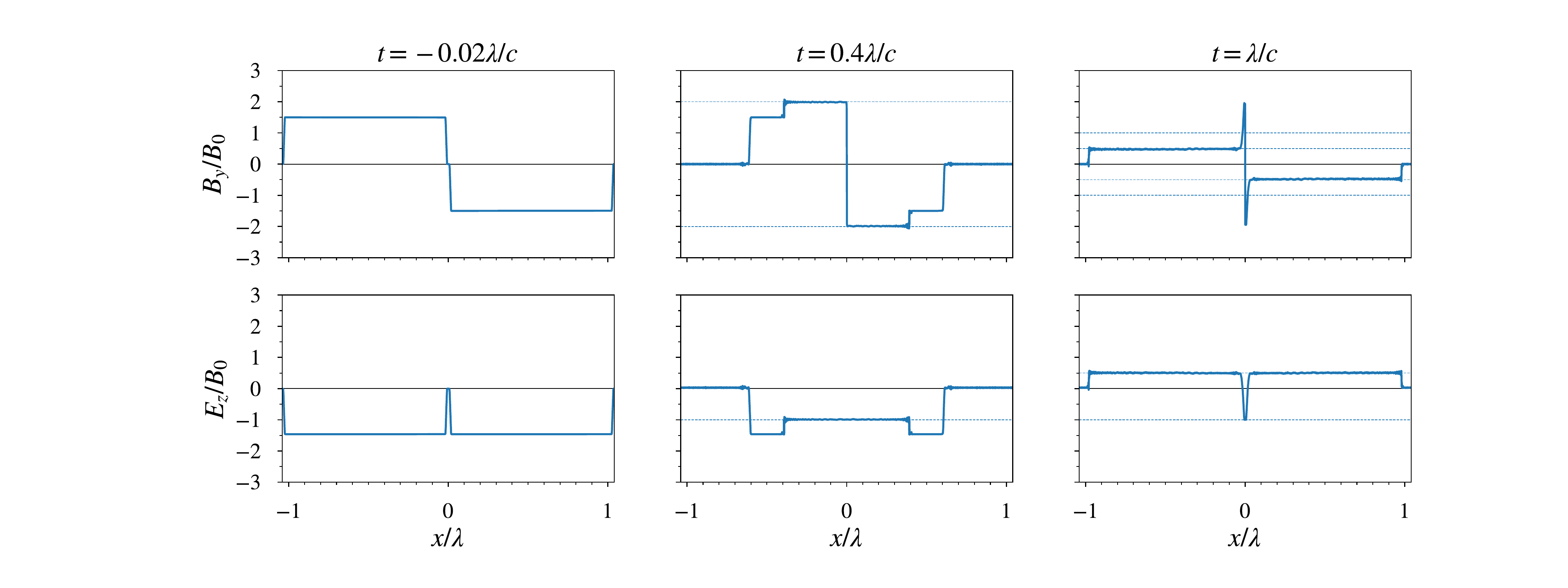}
\caption{Three snapshots of the 1D simulation with $\sigma_0=20$ and the initial amplitudes of the colliding packets $A=1.5$. 
The wave evolution agrees with the expected picture
shown in Figure~\ref{fig:squarewave}.}
\label{fig:snapshots}
\end{figure*}

\begin{figure}[t]
\includegraphics[width=0.45\textwidth]{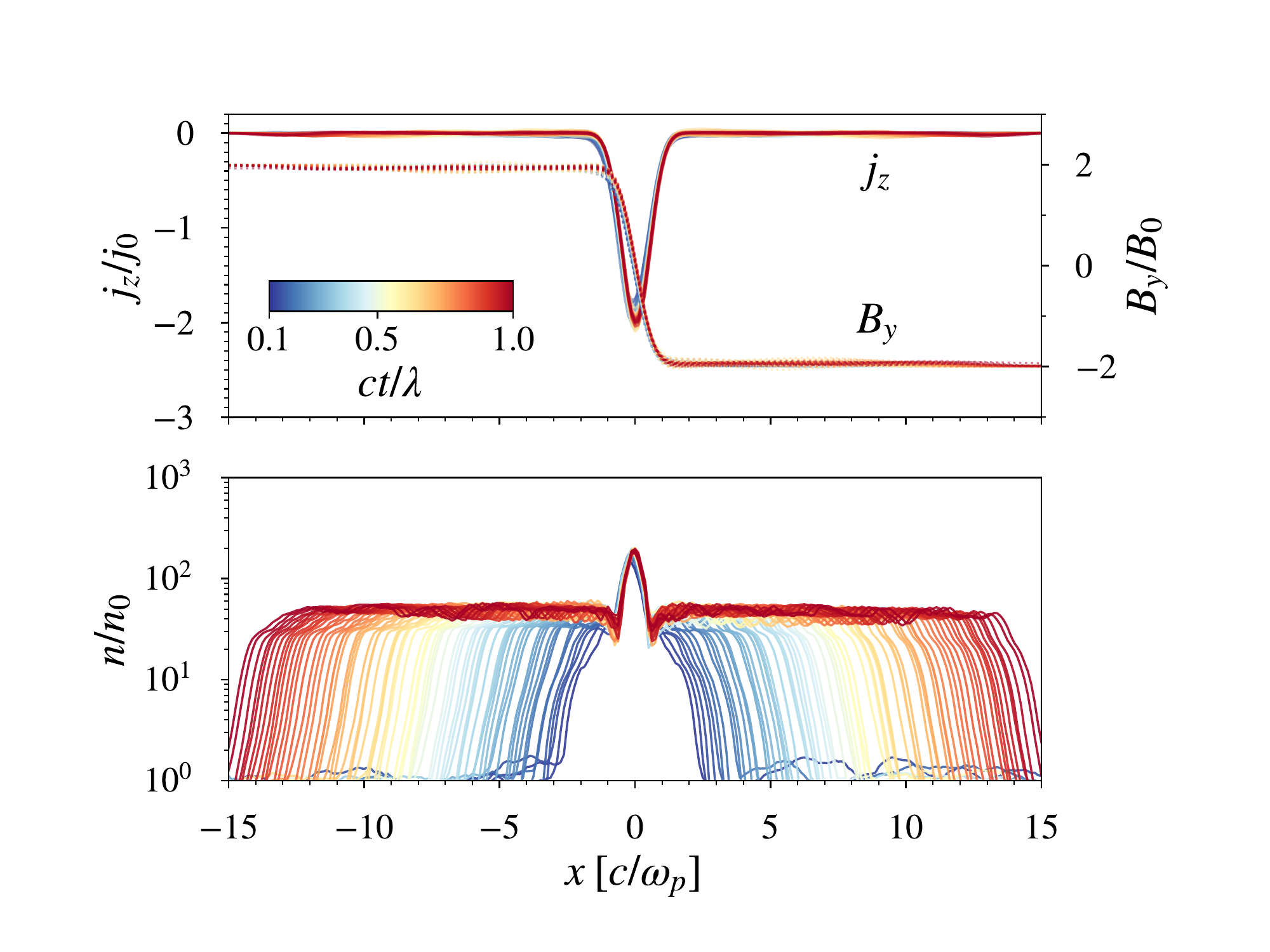}
\caption{Zoom-in of the region around the collision center $x=0$ from the simulation shown in Figure~\ref{fig:snapshots} ($\sigma_0=20$ and $A=1.5$).
Top: current density $j_z(x)$ in units of $j_0\equiv AB_0\omega_p/4\pi$ and magnetic field $B_y(x)$ in units of $B_0$.
Bottom: plasma density $n$ normalized to the pre-collision density $n_0$.
Curves with different colors correspond to different times $t$ indicated in the inset in the top panel. 
The curves of $j_z(x)$ (and $B_y(x)$) are nearly identical at different times, demonstrating that the current sheet is in a steady state.
The evolving curves of $n(x)$ demonstrate the slow, quasi-steady spreading of the dense layer of particles that have passed through and escaped the current sheet.
}
\label{fig:current}
\end{figure}

\begin{figure}[t]
\includegraphics[width=0.45\textwidth]{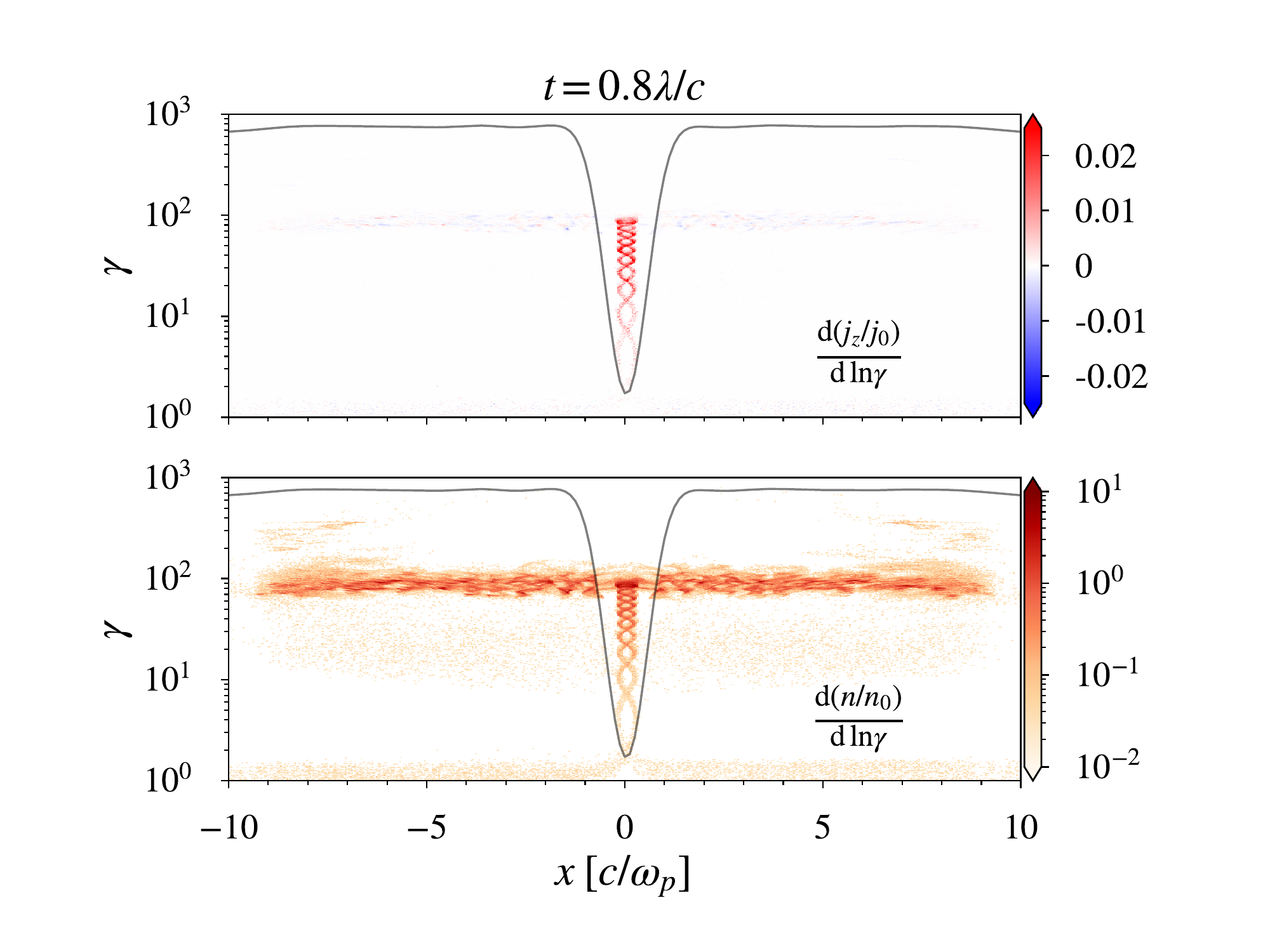}
\caption{Bottom: snapshot of the particle distribution in phase space (energy and position) at $t=0.8\lambda/c$, from the simulation with $\sigma_0=20$ and $A=1.5$. 
Color shows $d(n/n_0)/d\ln\gamma$ as indicated by the color bar.
Top: A similar plot for $d(j_z/j_0)/d\ln\gamma$. 
The black curve in each panel shows $j_z(x)$ (in arbitrary units), indicating the current sheet location.
}
\label{fig:mushroom}
\end{figure}

\begin{figure}[t]
\includegraphics[width=0.45\textwidth]{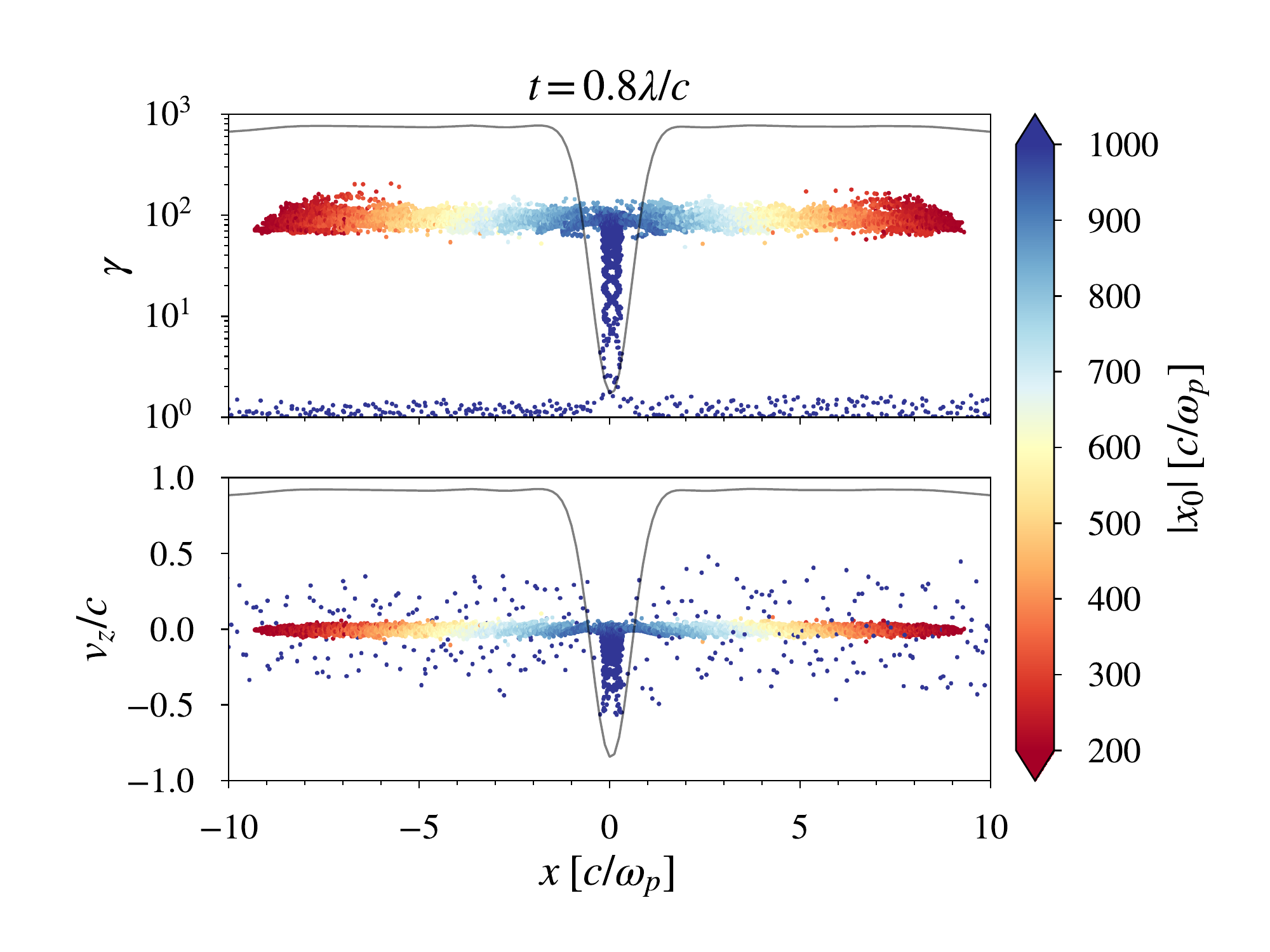}
\caption{
Top: snapshot of a particle population in phase space (energy and position) at $t=0.8\lambda/c$, from the simulation with $\sigma_0=20$ and $A=1.5$. 
The particles (positrons) were drawn according to their initial position $x_0$ at the beginning of the simulation; the particles were sampled uniformly in $|x_0|$ and traced throughout the simulation. 
The color code for $|x_0|$ is shown next to the figure; red particles are initially close to the center, and blue particles are initially far from the center, 
Bottom: the same set of particles is shown on the $x$-$v_z$ plane, at the same time $t=0.8\lambda/c$. The concentration of positrons with $v_z<0$ near $x=0$ sustains the electric current sheet.
The black curve in each panel shows $j_z(x)$ (in arbitrary units), indicating the current sheet location.
}
\label{fig:phase}
\end{figure}

Figures~\ref{fig:snapshots}-\ref{fig:phase} show the results of a sample simulation with $\sigma_0=20$ and $A=1.5$. 
The results are consistent with the analytical description in Section~\ref{sec:theory}. 
We observe that the collision launches two symmetric reflected waves with amplitudes $A-1=0.5$ (Figure~\ref{fig:snapshots}) and a quasi-steady current sheet is sustained at $x=0$ with width $\Delta\sim c/\omega_p$ (Figure~\ref{fig:current}).

To examine the plasma dynamics during the collision process, we traced a large sample of individual particles, starting from their initial positions $x_0$ at the beginning of the simulation and till the end of the packet interaction. 
Particles in the wave packet initially drift toward $x=0$ with speed $|\vdx^0|=c\,A^2/(A^2+1)\approx 0.7c$. 
They also have the $y$-component of the drift motion $v_{{\rm d},y}^0=cE_zB_x/B^2=\vdx^0/A\approx 0.46c$, and their initial Lorentz factor is $\gamma_{\rm d}^0=(A^2+1)^{1/2}\approx 1.8$.
When the particle encounters the reflected wave, its drift speed drops to $\vdx\approx 0.4$ according to Equation~(\ref{eq:vdx}).

When the drifting particle arrives to the collision center and enters the current sheet, it gains a high Lorentz factor over the timescale $\delta t$ given in Equation~(\ref{eq:delta_t}), and then exits the current sheet.
We observe that the particles exit with Lorentz factors $\sim 80$, in agreement with the analytical result $\bar{\gamma}\approx 82$ from Equation~(\ref{eq:gam_av}). 

After exiting the current sheet, the accelerated particles continue to gyrate about the magnetic field lines and also slowly slide along $\bB$ away from $x=0$. 
We observe that the accelerated particles form a slowly spreading cloud with density $n_{\rm acc}\sim 60 n_0$ (Figure~\ref{fig:current}), close to the analytic estimate $n_{\rm acc}\sim\bar{\gamma}n_0$ (Equation~\ref{eq:nacc}). 
The acceleration inside the current sheet followed by spreading with $\gamma\approx const$ outside the sheet creates the characteristic mushroom structure in the phase space, see Figures~\ref{fig:mushroom} and \ref{fig:phase}.

Particle acceleration increases their Larmor radii, which breaks the adiabatic ($\bE\times\bB$ drift) description of the particle motion inside the current sheet. 
As a result, the particles become capable of sustaining current $j_z$ along $\bE$ (as one can see in Figure~\ref{fig:mushroom}, the electric current is dominated by the energetic particles). 
This allows $j_z(x)$ to organize so that it self-consistently sustains the jump of $B_y(x)$ across the current sheet. 

The process of particle acceleration in the current sheet occurs in agreement with the description in Appendix~\ref{sec:app_motion}, which uses effective potential $V$ in the boosted frame $K'$. 
When viewed in the lab frame, the particle acceleration may be described similarly to \citet{1965JGR....70.4219S}.
The electrons and positrons are pushed by $E_z$ in the opposite $\pm z$ directions while the $x$-component of the Lorentz force $\pm e v_zB_y/c$ acts to confine the particles near $x=0$ between the opposite $B_y$. 
The Lorentz force also causes gyration in the $zy$ plane, as $\boldsymbol{v}$ rotates about $\bB_0$. 
This rotation occurs on the timescale $\delta t$, converting $v_z$ to $v_y$.
Once this rotation changes the sign of $v_z$, the $x$-component of the Lorentz force no longer confines the particles and they escape with the acquired energy.

\begin{figure}
\includegraphics[width=0.45\textwidth]{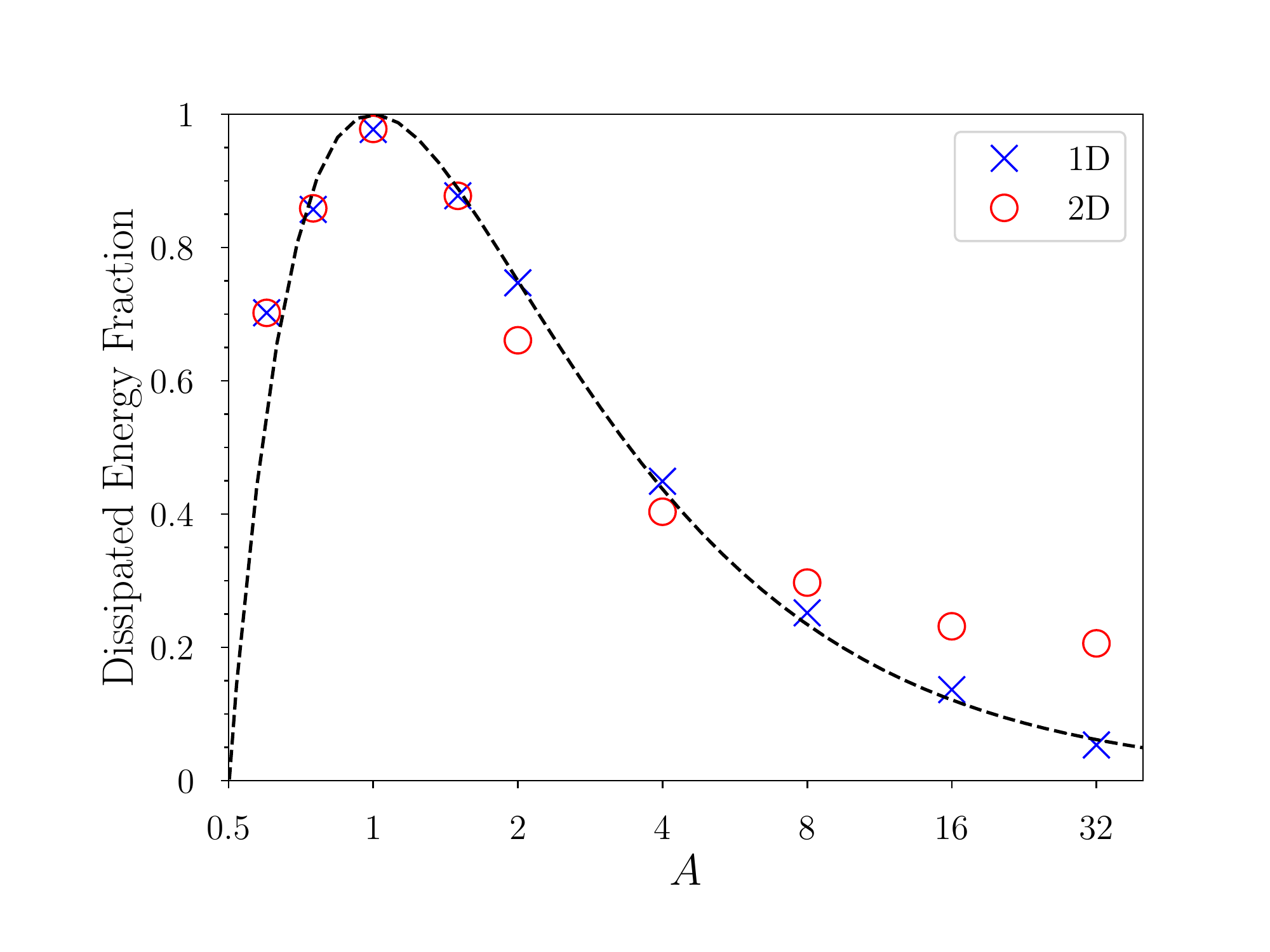}
\caption{Dissipated fraction of the wave energy, measured in the 1D and 2D simulations, as a function of the wave amplitude $A$ (the magnetization parameter $\sigma_0=20$ in all the simulations). 
The 1D simulations (blue crosses) reproduce the analytical result $f=(2A-1)/A^2$ obtained in the limit of $\sigma_0\gg 1$ (dashed curve).
The 2D results (red circles) deviate from the 1D results at $A\gtrsim 2$, because the current sheet at the collision center becomes tearing unstable.
}
\label{fig:energy}
\end{figure}

The dissipated fraction of the packet energy observed in the fiducial simulation with $A=3/2$ agrees with $f=8/9$ predicted by Equation~(\ref{eqn:frac_dis}). 
We have also performed eight other simulations with different packet amplitudes $A$. 
The results reproduced the predicted dependence $f(A)=(2A-1)/A^2$, as shown in Figure~\ref{fig:energy}.

\subsection{2D Simulations}\label{sec:sim2d}

We have run a set of 2D models with $\sigma_0=20$ and various values of $A$, from $A=0.6$ to $A=32$.
The 2D computational box is in the $xy$ plane, with the retained translational symmetry along $z$. 
In most simulations, the number of cells in the $x$ direction is $N_x=51200$, the same as in the 1D model, and we use $N_y=640$ cells in the $y$ direction, with periodic boundary conditions.
We have also performed simulations with $N_x=25600$ and $N_y=2560$. The large $N_y$ becomes important if the current sheet at the collision interface develops a tearing instability, which occurs at large $A$. 
We used $N_x\times N_y=25600\times 2560$ for models with $A\geq 8$, keeping the same condition of 8 cells per skin depth $c/\omega_p$. 
In these models, the reduced box size in the $x$ direction requires the reduction of the packet length by a factor of 2, from $\lambda=3072\,c/\omega_p$ to $\lambda=1536\,c/\omega_p$. 

The 2D models with $A<2$ show almost exactly the same evolution as the corresponding 1D models. 
This may be expected, as the cross-layer magnetic field $B_x=B_0$ tends to stabilize the current sheet against the tearing instability \citep{1976JETP...43.1113G}. 
We observed that the colliding wave packets are partially reflected and partially absorbed by the current sheet, exactly as in the 1D model, with the same dependence of the absorbed fraction $f$ on the wave amplitude $A$ (Figure~\ref{fig:energy}). 

\begin{figure}[t]
\includegraphics[width=0.45\textwidth]{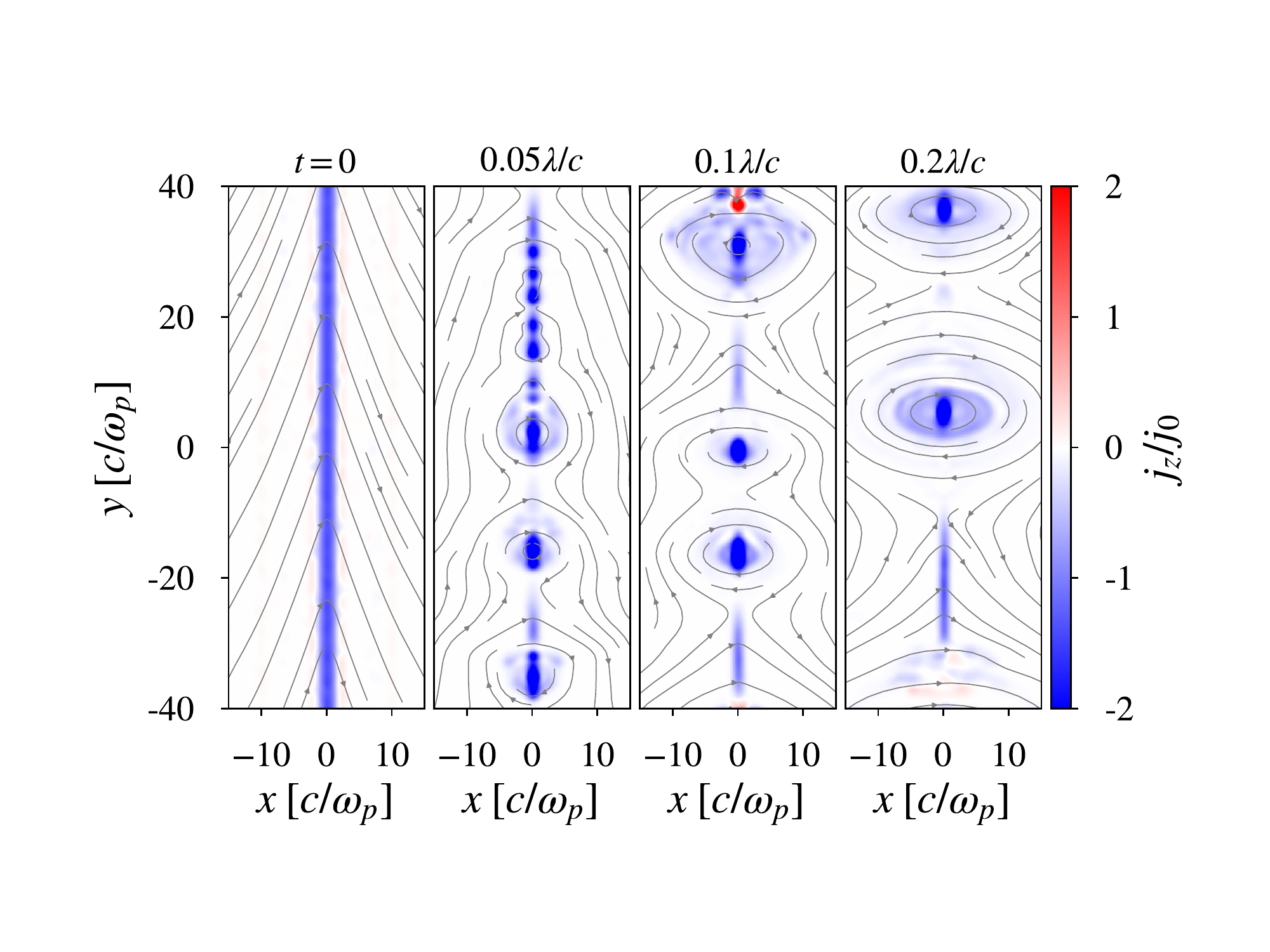}
\caption{Snapshots of current density $j_z(x,y)$ in the 2D simulation with $A=2$ and $\sigma_0=20$. 
The figure shows the region near the collision centre $x=0$ where the current sheet forms. 
The snapshots were taken at times $t$ indicated at the top of each panel.
One can see the development of tearing instability of the current sheet.}
\label{fig:current2d}
\end{figure}

\begin{figure}[t]
\includegraphics[width=0.45\textwidth]{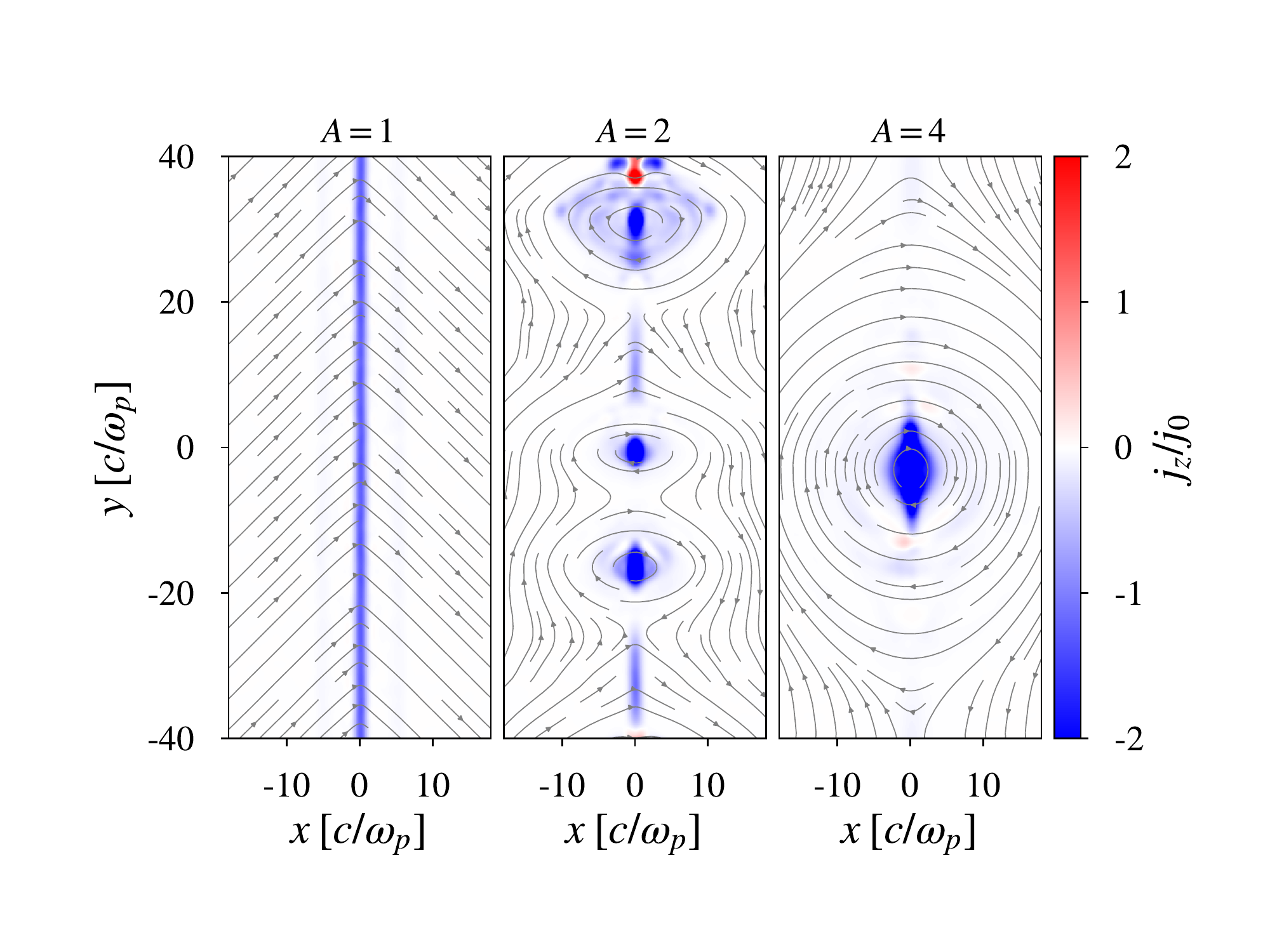}
\caption{Snapshots of current density $j_z(x,y)$ taken at $t=0.1\lambda/c$ in three 2D simulations with $\sigma_0=20$, which have different amplitudes of the wave packets $A=1$, 2, 4 (as indicated at the top of each panel). 
Tearing instability and magnetic reconnection did not occur in the simulation with $A=1$. 
At larger $A$, magnetic reconnection generated multiple plasmoids, which merged into bigger plasmoids over time, in particular in the simulation with $A=4$. }
\label{fig:current2dcomp}
\end{figure}

The stabilization effect of $B_0$ disappears at sufficiently small $B_0$ or, equivalently, at sufficiently large $A$. 
We observed that in the models with $A\geq 2$, the plane symmetry of the current sheet becomes broken by the tearing instability, initiating the process of magnetic reconnection.
The development of tearing instability in the model with $A=2$ is shown in Figure~\ref{fig:current2d}.
Tearing triggers magnetic reconnection that converts $B_y$ into $B_x$ and produces closed magnetic islands  (``plasmoids'') in the $xy$ plane. 
The plasmoids begin to move along the current sheet and merge.
This process is activated at $A\sim 2$ and fully develops at $A>2$, as demonstrated in Figure~\ref{fig:current2dcomp} by comparing the current sheet in the models with $A=1$, 2, and 4 at the same time $t=0.1\lambda/c$.
One can see that plasmoid merging has already produced a monster plasmoid in the $A=4$ simulation. 
The merging process is slower in the model with $A=2$, and there is no plasmoid formation at $A=1$. 

\begin{figure*}[t]
\includegraphics[width=\textwidth]{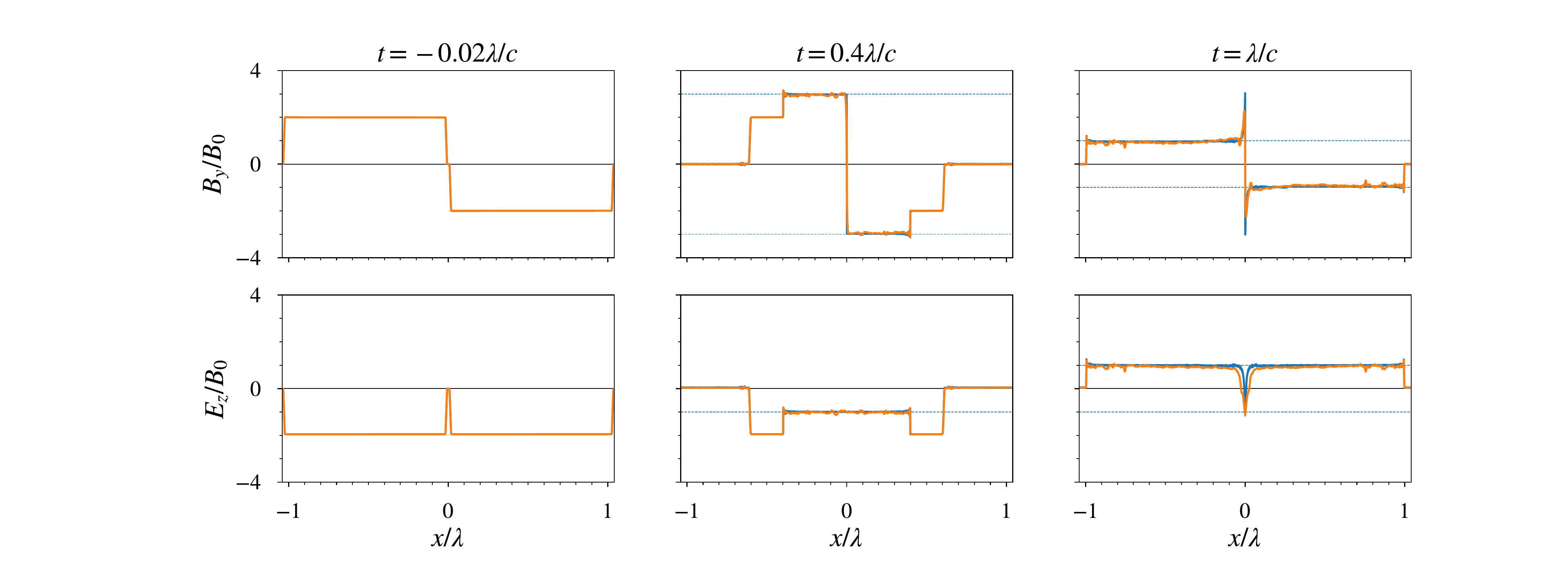}
\caption{Three snapshots of the wave fields $B_y(x)$ and $E_z(x)$ in the 1D (blue) and 2D (orange) simulations with $A=2$ and $\sigma_0=20$. 
In the 2D model, $B_y(x)$ and $E_z(x)$ were obtained by averaging along the $y$ direction. 
The $x$ profiles found in the 1D and 2D models are nearly identical despite the fact that $A=2$ is sufficiently large to allow tearing of the current sheet in the 2D simulation.  
}
\label{fig:snapshots2d}
\end{figure*}

In the models with $A=2, 4, 8$, the onset of magnetic reconnection has a modest effect on the wave reflection by the current sheet. 
As a result, the dissipation fraction $f$ in these models does not strongly deviate from the prediction of the 1D model (Figure~\ref{fig:energy}).
Figure~\ref{fig:snapshots2d} compares the $x$ profiles of the electromagnetic fields in the 1D and 2D simulations with $A=2$ (in the 2D model, the $x$ profile was obtained by averaging the field values along the $y$ direction).
One can see that the profiles are nearly the same in the 1D and 2D simulations despite the fact that the current sheet experiences the tearing instability and breaks up into magnetic plasmoids (Figure~\ref{fig:current2dcomp}).

At yet larger wave amplitudes, $A=16$ and $32$, the 1D and 2D results strongly differ. 
The 1D model predicts a small dissipation efficiency $f\approx 2/A$ at $A\gg 1$. 
The 2D simulations demonstrate that magnetic reconnection due to tearing instability becomes the dominant form of dissipation, releasing $\sim 0.2$ of the wave energy. 
For instance, in the 2D model with $A=32$, the dissipated fraction $f$ exceeds the prediction of the 1D model by a factor of $\sim 4$. 

The value of the dissipated energy fraction in the $A\gg1$ limit can be understood as follows. 
Near the reconnection layer, the magnetic field amplitude is $|B_y|\approx 2 A B_0$.
For a reconnection rate $\eta\sim 0.1$, the Poynting flux converging into the layer from each side is $S_{\rm rec}\simeq \eta c (2 A B_0)^2/4 \pi$.
Recent studies of relativistic reconnection show that roughly half of the electromagnetic energy advected into the layer converts to plasma energy \citep[e.g.,][]{sironi_15}, so the dissipated energy flux is $\sim S_{\rm rec}/2$. 
This implies that the colliding waves are dissipated with efficiency 
\begin{equation}
   f\sim \frac{S_{\rm rec}}{2S_0}=2\eta\sim 0.2 \qquad (A\gg 1).    
\end{equation}
The reconnection process will continue as long as the incoming and reflected waves overlap near the layer, i.e., for a time of $\lambda/c$.

\section{Discussion}\label{sec:discussion}
Alfv\'en waves in a strongly magnetized plasma ($\sigma_0\gg 1$) have a special feature: the wave propagates with nearly speed of light $c$, which implies nearly equal electric and magnetic fields in the wave, $\delta E=\delta B$. 
For strong waves, $A=\delta B/B_0\sim 1$, this opens the possibility for $E=\delta E$ to exceed the magnetic field $B=\sqrt{(\delta B)^2+B_0^2}$, triggering fast dissipation. 
This effect is most prominent in collisions of linearly polarized Alfv\'en waves with anti-aligned $\delta \bB$ and aligned $\delta\bE$. 
We have investigated the collision of two symmetric plane waves propagating with wave vectors $\pm \boldsymbol{k}\parallel \bB_0$ in a uniform background field $\bB_0$. 
In this case, the problem becomes essentially one-dimensional, i.e. all quantities depend on the $x$ coordinate that runs along $\bB_0$.
We have described the collision analytically and performed kinetic simulations, both 1D and 2D, allowing spontaneous breaking of the plane symmetry. 
Our fiducial numerical model has $\sigma_0=20$ and $A=1.5$, and we also calculated a broad range of models with $\sigma_0\gg 1$ and $A\leq 32$. 
Our results are as follows.
 
\begin{enumerate}
\item
Dissipation is triggered when the colliding waves have the amplitudes $A>1/2$. Waves with $A<1/2$ pass through each other without dissipation.
\item
The collision with $A>1/2$ generates a thin current sheet at the collision center $x=0$ which separates the opposite magnetic fields of the colliding waves. 
The sheet serves as a wall that partially reflects and partially absorbs the incoming waves. 
It self-organizes to sustain $E=B_0$, so that $E=B$ at $x=0$. 
This condition determines the reflected wave amplitude $A_r=A-1$ and the absorbed energy fraction $f=(2A-1)/A^2$. 
When $A=1$ there is no reflection and $100\%$ of the wave energy is dissipated, i.e. the two waves annihilate each other at the collision center.
\item
The dissipation generates particles with Lorentz factors $\bar{\gamma}\sim A^2\sigma_0$ (Equation~\ref{eq:gam_av}). 
The energy conversion process is quasi-steady when the length of the colliding wave packets $\lambda\gg \bar{\gamma}\, \Delta$ where $\Delta$ is the thickness of the current sheet.  
\item 
The current sheet thickness is self-regulated to $\Delta \sim c/\omega_p$, where $\omega_p^2=4\pi e^2 n_0/m_e$ is defined for the plasma in the wave packet before the collision. 
The colliding waves feed particles into the current sheet with a mildly relativistic drift speed. 
The particles enter and then exit the sheet, gaining the Lorentz factor $\bar{\gamma}$  on the timescale $\delta t\sim \bar{\gamma}^{3/2}/\omega_B$, where $\omega_B=eB_0/mc$.
The energized particles gyrate about $\bB$ and spread from the sheet (upstream into the incoming wave packet) along $\bB$ with a small $v_\parallel\sim \pm c/\bar{\gamma}$. 
The spreading cloud of accelerated particles has density $n_{\rm acc}\sim \bar{\gamma} n_0$, much higher than the original plasma density $n_0$.
\item 
The current sheet becomes tearing unstable for sufficiently large $A$.
In the limit of $A\rightarrow\infty$ ($B_0\rightarrow 0$) the packet collision setup becomes the standard configuration for reconnection of $\pm B_y$, with no cross-layer field $B_x$. 
Then, energy release is driven by the tearing instability. 
It proceeds with a smaller rate $v_{\rm rec}\sim 0.1 c$ (instead of $c$) and the energized particles are ejected in plasmoids along $\pm y$, i.e. along the current sheet (instead of spreading into the upstream along $\pm x$). 
Reconnection dissipates $\sim 0.2$ of the energy of the colliding waves.
\end{enumerate}

The collision of Alfv\'en waves with amplitudes $A>1/2$ in a plasma with $\sigma_0\gg 1$ provides an extremely fast and efficient dissipation mechanism. 
In particular, for $A=1$ the entire wave energy converts to heat in one light-crossing time of the packet, $\lambda/c$. 
This mechanism can operate in strongly perturbed magnetized plasmas around compact objects, including jets from black holes, coronae of accretion disks, magnetospheres of neutron stars, and pulsar winds.
It occurs significantly faster than magnetic reconnection or
the turbulence cascade to a small dissipation scale.

Our kinetic simulations and the analytical description of wave collision neglected radiative losses of the accelerated particles. 
The losses are unlikely to reduce the dissipation efficiency or change the collision picture with the current sheet, but can change the particle trajectories in the sheet. 
In particular, for waves colliding in a neutron star magnetosphere, synchrotron losses will limit the excitation of ultra-relativistic gyration in the current sheet. 
Furthermore, Compton scattering will generate energetic photons which convert to $e^\pm$ pairs. 
These radiative processes will occur similarly to those in radiative magnetic reconnection \citep{2020arXiv201107310B} and generate a powerful X-ray burst with a similar spectrum. 

Alfv\'en waves excited at larger distances from the compact object (e.g. in a magnetized jet from a black hole or in a pulsar wind nebula) may dissipate with less severe radiative losses and with no $e^\pm$ creation.
They can still produce observable gamma-ray flares, similar to magnetic reconnection invoked previously to explain blazar flares \citep[e.g.,][]{2009MNRAS.395L..29G}.

\bigskip 

\acknowledgements
X.L. is supported by the Natural Sciences and Engineering Research Council of Canada (NSERC), funding reference \#CITA 490888-16 and acknowledges the support of computational resources provided by Compute Ontario and Compute Canada.
Research at Perimeter Institute is supported in part by the Government of Canada through the Department of Innovation, Science and Economic Development Canada and by the Province of Ontario through the Ministry of Colleges and Universities.
A.M.B. is supported by NASA grant NNX~17AK37G, NSF grant AST~2009453, Simons Foundation grant \#446228, and the Humboldt Foundation.
L.S. acknowledges support from the Sloan Fellowship, the Cottrell Scholar Award, NASA 80NSSC20K1556, NSF PHY-1903412 and DoE DE-SC0021254. 

\appendix

\section{Particle Motion in Uniform Perpendicular Electromagnetic Fields}\label{sec:app_motion1}
A test particle with mass $m$ and charge $q$ moving at velocity $\boldsymbol{v}$ in an electromagnetic field feels the Lorentz force
\begin{equation}
    \frac{\md }{\md t} (\gamma m\boldsymbol{v}) = q\left(\be + \frac{\boldsymbol{v}}{c} \times \bb\right).
\end{equation}
Since the magnetic force $\boldsymbol{v}\times\bb/c$ does no work on the particle, the change of particle energy arises only from the electric force
\begin{equation}
    \frac{\md \gamma}{\md t} = \frac{q}{m c^2} \be\cdot\boldsymbol{v}.
\end{equation}
When the electromagnetic field is constant in time, analytical solution can be obtained by changing the reference frame so that only the electric or magnetic field exists.

\begin{enumerate}
    
\item $B>E$.
    
If the magnetic field dominates, the electric field vanishes in a reference frame which is moving with drift velocity
\begin{equation}
    \boldsymbol{v}_d = \frac{\be\times\bb}{B^2}c.
\end{equation}
In that reference frame, $\bE^\prime=0$, and the test particle will gyrate around the constant magnetic field $\bB^\prime\parallel \bB$.
Therefore, the particle's motion viewed in the lab frame is a combination of gyration around $\bb$ and drift with $\boldsymbol{v}_d$.
The particle energy in the lab frame oscillates with the  gyration period
with no systematic energy gain on longer timescales, as the time-averaged $\overline{\be\cdot\boldsymbol{v}}=0$.

\item $E>B$.

If the electric field dominates, the magnetic field vanishes in the  frame moving with the velocity
\begin{equation}
     \boldsymbol{v}_d = \frac{\be\times\bb}{E^2}c.
\end{equation}
In this new frame, $\bB^\prime=0$, and the particle is linearly accelerated by $\bE^\prime\parallel \bE$. 
The resulting motion in the lab frame will be a combination of linear acceleration along $\be$ and constant motion with $\boldsymbol{v}_d$.
The particle is continually gaining energy as $\be\cdot\boldsymbol{v}$ is always positive. 

\item $E=B$.

In the special case of $E=B$, there is no  reference frame where only $E$ or $B$ is non-zero. Since $E^2-B^2$ is a Lorentz invariant, $E=B$ in all frames.
The analytical solution for particle motion in this special case is given in \citet{landau1975classical}.
When the magnetic field is pointing in the $y$ direction and the electric field in the $z$ direction, there are two constants of motion which are the $y$-momentum $p_y=\gamma m v_y$ and $\alpha\equiv \gamma m c^2(1-v_x/c)$, where $\gamma=(1-v^2/c^2)^{-1/2}$.
The particle acceleration along $z$ direction is described by equation
\begin{equation}
    2eEt = \left( 1+\frac{\varepsilon^2}{\alpha^2} \right)p_z +\frac{c^2}{3\alpha^2}p_z^3,
\end{equation}
where $\varepsilon^2=m^2c^4+p_x^2 c^2$.
The $x$-momentum of the particle grows fastest (the $-x$ axis is along $\boldsymbol{E}\times\boldsymbol{B}$)
\begin{equation}
    p_x = -\frac{\alpha}{2c} + \frac{p_z^2c^2+\varepsilon^2}{2\alpha c}.
\end{equation}
The particle energy is given by 
\begin{equation}
    \gamma mc^2 = \frac{\alpha}{2} + \frac{p_z^2c^2+\varepsilon^2}{2\alpha}.
\end{equation}

\end{enumerate}


\section{Particle Motion in a Static Magnetic Jump with electric Field}\label{sec:app_motion}
Let us consider the motion of a positron in a static electromagnetic field $\bB=(B_0,B_y(x),0)$ and $\bE=(0,0,E_z)$, where 
\begin{equation}
 E_z=-aB_0, \qquad B_y=bB_0 h(x).
\end{equation}
Here $a>0$ and $b>0$ are constants, and $h(x)$ is a smooth function monotonically decreasing from $h\approx 1$ at $x<- \Delta$ to $h\approx -1$ at $x>\Delta$. 
This electromagnetic configuration describes the jump of $B_y(x)$ on the scale $\Delta$ that forms during the collision of two symmetric \alfven packets, as observed in our simulations (with $b=2A-1$, see Equation~(\ref{eq:By})).
Non-relativistic particle motion in a magnetic jump with $h(x)=-x/\Delta$ at $|x|<\Delta$ was studied by \citet{1965JGR....70.4219S}. 
We are interested here in the relativistic case where the particle is accelerated by $E_z$ to a high Lorentz factor $\gamma$.
The constant $a$ is smaller than unity but close to it.

The particle motion obeys the equation $d\boldsymbol{p}/dt=e(\boldsymbol{v}\times\bB/c+\bE)$, where $\boldsymbol{p}=\gamma m\boldsymbol{v}$ and $\boldsymbol{v}$ is the particle velocity. 
Since $a<1$, $E_z$ can be removed by a Lorentz boost to frame $K^\prime$ moving with velocity $v_F=ac$ along $-y$; the frame Lorentz factor is $\gamma_F=(1-a^2)^{-1/2}$.
We will denote all quantities measured in frame $K^\prime$ with a prime, so
\begin{equation}
    \bE^\prime=0, \qquad \boldsymbol{B}' = (B'_x,B'_y,0) = \left(\frac{B_0}{\gamma_F},B_y,0\right),  
 \qquad {B^\prime}^2=B^2-E^2=(1+b^2h^2-a^2)B_0^2.
\end{equation}
The equation of motion in frame $K^\prime$ becomes 
\begin{equation}
\label{eq:dpdt}
    \frac{d\boldsymbol{p}^\prime}{dt^\prime}=\frac{e}{c}\, \boldsymbol{v}^\prime\times\bB^\prime.
\end{equation}
Note that the boost to frame $K^\prime$ leaves $x^\prime=x$ and $B_y=B_y^\prime$ the same, so it does not affect the magnetic jump profile, $B_y^\prime(x^\prime)=B_y(x)$. 

Let us consider a particle moving at $|x^\prime|=|x|>\Delta$ (where $|h|\approx 1$), and approaching the magnetic jump with the drift velocity  in the lab frame,
\begin{equation} 
   \boldsymbol{\beta}_d=\frac{\boldsymbol{v}_d}{c}=\left(-\frac{E_zB_y}{B^2},\frac{E_zB_x}{B^2},0\right)
   =\left(\frac{ab}{1+b^2},-\frac{a}{1+b^2},0\right).
\end{equation}
The drift Lorentz factor in the lab frame is given by
\begin{equation}
    \gamma_d=\frac{1}{\sqrt{1-E^2/B^2}}
    =\sqrt{\frac{1+b^2}{1+b^2-a^2}}
    \approx \sqrt{1+\frac{1}{b^2}}.
\end{equation}
In frame $K^\prime$, the particle is initially moving along $\bB^\prime$. 
Its Lorentz factor and velocity are given by the transformation of the drift four-velocity $(\gamma_dc,\gamma_d\boldsymbol{v}_d)$,
\begin{equation}
\label{eq:v0prime}
   \gamma^\prime=\gamma_F\gamma_d(1+\beta_F\beta_{d,y})=\frac{\gamma_F}{\gamma_d}, \qquad
   \frac{\boldsymbol{v}_0^\prime}{c}=\left(\frac{ab}{\gamma_F(1+b^2-a^2)},\frac{ab^2}{1+b^2-a^2},0\right)\approx \left(\frac{a}{\gamma_Fb},a,0\right).
\end{equation}
Since $\bE^\prime=0$, $\gamma^\prime=\gamma^\prime_0$ remains constant even after the particle enters the magnetic jump while its velocity vector $\boldsymbol{v}^\prime$ will rotate from $\boldsymbol{v}_0^\prime$ keeping $|\boldsymbol{v}^\prime|=|\boldsymbol{v}_0^\prime|$.
Note that $v_{y0}^\prime\approx c$ and $v_{x0}^\prime=(B_x^\prime/B^\prime)c\ll c$. 
After entering the region $|x^\prime|=|x|<\Delta$, where the magnetic field lines are curved, the particle develops a significant velocity component perpendicular to $\bB^\prime$ and begins to gyrate about $\bB^\prime$. 

\begin{figure}
\centering
\includegraphics[width=0.45\textwidth]{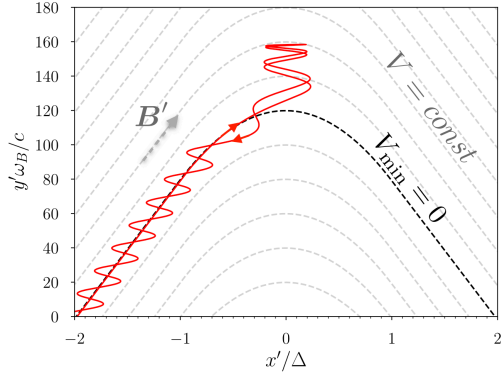}
\caption{Particle trajectory (red curve)
in frame $K'$, calculated for a linear magnetic jump $B_y'(x')=B_y(x)=-bB_0x/\Delta$. In this example $b=2$, $\sigma_0=20$, and $\Delta=c/\omega_p$. 
Then the final Lorentz factor of the particle exiting the current sheet in the lab frame is $\gamma=82$ (Equation~\ref{eq:gam_av}), and the corresponding Lorentz factor in frame $K'$ is $\gamma'=\sqrt{\gamma}\approx 9=const$. 
The grey dashed curves show the magnetic field lines, which also represent the isocontours of effective potential $V(x',y')$ that governs the particle motion (Equation~\ref{eq:V}). 
The particle initially moves along the straight line at the bottom of the potential valley $V_{\min}=0$ (black dashed curve); this initial motion corresponds to the $\bE\times\bB$ drift in the lab frame. When reaching the sharp turn of the valley (the magnetic jump) the particle inertia makes it climb the concave side of the valley, then descend back to the bottom and exit, with oscillations.
}
\label{fig:trajectory}
\end{figure}

It is convenient to describe the magnetic field $\bB^\prime=\nabla\times \boldsymbol {A}^\prime$ using vector potential $A_z^\prime(x^\prime,y^\prime)$: $B_x^\prime=\partial A_z^\prime/\partial y^\prime$ and $B_y^\prime=-\partial A_z^\prime/\partial x^\prime$. 
Note that $A_z^\prime=const$ along the magnetic field lines, $\bB^\prime\cdot\nabla A_z^\prime=0$. 
Symmetry $\partial/\partial z^\prime=0$ implies conservation of the generalized $z$-momentum $\gamma^\prime m v_z^\prime+(e/c)A_z^\prime=const$. 
This gives
\begin{equation}
\label{eq:vz}
    v_z^\prime=-\frac{e}{\gamma^\prime mc}\,A_z^\prime(x^\prime,y^\prime),
\end{equation}
where we chose $A_z^\prime=0$ for the magnetic field line that is followed by the particle before it enters the magnetic jump (so that $v_z^\prime=0$ corresponds to $A_z^\prime=0$). 
Equation~(\ref{eq:vz}) is the integral of the $z$-component of the dynamic Equation~(\ref{eq:dpdt}).
Using  ${v_x^\prime}^2+{v_y^\prime}^2+{v_z^\prime}^2={v_0^\prime}^2=const$ ($\gamma^\prime=const$), one finds
\begin{equation}
\label{eq:V}
    \frac{{v_x^\prime}^2}{2}+\frac{{v_y^\prime}^2}{2}+V(x^\prime,y^\prime)=const=\frac{{v_0^\prime}^2}{2},
    \qquad V(x^\prime,y^\prime)=\frac{1}{2}\left(\frac{eA_z^\prime}{\gamma^\prime mc}\right)^2.
\end{equation}
The $x^\prime$, $y^\prime$ components of the dynamic equation~(\ref{eq:dpdt}) can now be written in the form,
\begin{equation}
    \frac{dv_x^\prime}{dt^\prime}=-\frac{\partial V}{\partial x^\prime}, 
    \qquad 
    \frac{dv_y^\prime}{dt^\prime}=-\frac{\partial V}{\partial y^\prime}.
\end{equation}
The particle motion in the $x^\prime y^\prime$ plane is equivalent to non-relativistic dynamics in the given potential $V(x^\prime,y^\prime)$. 
This gives an intuitively clear picture of the particle motion.
Before entering the region $|x^\prime|<\Delta$, the particle moves along the line of $V=V_{\min}=0$, at the bottom of the potential valley. 
In the region $|x^\prime|<\Delta$ the line of $V=V_{\min}$ (the magnetic field line $y^\prime_{\min}(x^\prime)$) curves away from the straight line, and inertia makes the particle deviate from $y^\prime_{\min}(x^\prime)$ and ballistically climb the concave side of the valley up to some $V_{\max}>0$. 
Then the particle slides back to the bottom and exits to $|x^\prime|>\Delta$ (Figure~\ref{fig:trajectory}). 
The shape of the curved valley is displayed by the equipotentials $V=const$, i.e. by the magnetic field lines, which change direction by nearly $180^\circ$ at $|x^\prime|<\Delta$, where $B_y^\prime$ changes from $bB_0\gg B_x^\prime$ to $-bB_0$.
The concave shape of the valley's side with positive $\delta y^\prime=y^\prime-y^\prime_{\min}(x^\prime)$ results in $x^\prime$  oscillations about $0$ while the particle is climbing the slope to $\delta y^\prime_{\max}$ and then descending back to $\delta y^\prime=0$.
After exiting the curved region $|x^\prime|<\Delta$ the particle moves along the straight valley with some velocity $v_\parallel^\prime$ (parallel to the line of $V=0$) and also oscillates around $V=0$ with velocity $\boldsymbol{v}^\prime_\perp\parallel \nabla V$ (this means that the particle gyrates about the magnetic field line). 
The sum ${v_\perp^\prime}^2+{v_\parallel^\prime}^2={v_0^\prime}^2\approx c^2$ remains unchanged. Note that the particle enters the curved valley region with positive $v_y^\prime\approx c$ and exits with $v_y^\prime\sim -c$. 
This implies a big change of the particle Lorentz factor in the lab frame $\gamma$, as seen from the Lorentz transformation,
\begin{equation}
\label{eq:gamma}
    \gamma=\gamma^\prime\gamma_F(1-\beta_y^\prime\beta_F).
\end{equation}
The initial $\gamma=\gamma_d$ corresponds to $\beta_y^\prime=v_y^\prime/c\approx \beta_F$. 
When $\beta_y^\prime$ changes sign, the transformation~(\ref{eq:gamma}) gives $\gamma\sim \gamma^\prime\gamma_F=\gamma_F^2/\gamma_d\approx \gamma_F^2$. 
This Lorentz factor represents the energy gain of the particle in the magnetic jump, which we evaluated as $\bar{\gamma}$ in Equation~(\ref{eq:gam_av}). 
Therefore, 
\begin{equation}
    \gamma^\prime\approx \gamma_F\approx \bar{\gamma}^{1/2}\propto \sigma_0^{1/2}.
\end{equation}

The characteristic duration of the strong deviation $\delta y^\prime=y^\prime-y^\prime_{\min}>0$ (during which the particle ascends and descends the concave side of the turning valley) is $\delta t^\prime_1\sim \gamma^\prime\gamma_F/\omega_B$. 
It is the timescale for gyration  about $B_x^\prime=B_0/\gamma_F$ that converts $v_y^\prime>0$ to $v_z^\prime<0$ (ascent) and then $v_z^\prime<0$ to $v_y^\prime<0$ (descent), allowing the particle to turn and exit the curved region of the valley. 
During this turning phase $v_z^\prime=-(e/mc)B_x^\prime \delta y^\prime$ reaches a maximum negative value $\approx -c$. 
Thus, during the time $\sim \delta t^\prime_1$ the particle makes a large contribution to the electric current $j_z=j_z^\prime<0$. 
Note that $j_z<0$ is needed to self-consistently sustain the magnetic jump $dB_y/dx<0$.

Another characteristic timescale $\delta t_2^\prime=\Delta/v_{x0}^\prime$ shows how long it would take the particle to cross the curved region with $v_{x0}^\prime\approx c/b\gamma_F$ (Equation~\ref{eq:v0prime}).  
The main dimensionless parameter of the magnetic jump is the ratio 
\begin{equation}
    Q\equiv\frac{\delta t_2^\prime}{\delta t_1^\prime}= \frac{b\omega_B\Delta}{\gamma^\prime c}, \qquad \omega_B=\frac{eB_0}{mc},
\end{equation}
which scales with width $\Delta$.
If $Q\gg 1$, the particle closely follows the curved magnetic field line in most of the region $|x^\prime|<\Delta$, with a small $\delta y^\prime=y^\prime-y^\prime_{\min}(x)$ and a small $|v_z^\prime|\ll c$; the particle quickly turns near $x^\prime= 0$. $Q\gg 1$ implies that particles entering and exiting the magnetic jump create a significant $j_z$ during a small fraction of their residence time at $|x^\prime|<\Delta$, and the resulting $j_z$ tends to be localized at $|x^\prime|\ll \Delta$. 
In this case, the plasma inside the magnetic jump fails to self-consistently sustain $4\pi j_z=c \,dB_y/dx$ at $|x^\prime|<\Delta$. 
In a self-consistent situation, the jump adjusts its width so that $Q\sim 1$. 
This fact is observed in our kinetic simulations.

The particle exits the region of $|x^\prime|<\Delta$ with $|\bar{v}_x^\prime|=v_\parallel^\prime B_x^\prime/B^\prime$ (averaged over gyrations), which is comparable to the initial $v_{x0}^\prime\approx c/b\gamma_F$. The corresponding $v_x$ in the lab frame is found from $\gamma v_x=\gamma^\prime v_x^\prime$, which gives 
\begin{equation}\label{eq:vx}
    |\bar{v}_x|\sim \frac{\gamma^\prime}{\bar{\gamma}}\,|\bar{v}_x^\prime|\sim \frac{c}{b\bar{\gamma}}. 
\end{equation}
Time $dt$ (measured in the lab frame) is related to $dt^\prime$ along the world line of the particle by the transformation $dt^\prime=\gamma_F(1-\beta_F\beta_y^\prime)dt^\prime$. 
At the entrance to the layer $|x|<\Delta$, this gives $dt=dt^\prime/\gamma_F$. 
However, after the particle loses its $\beta_{y0}^\prime\approx \beta_F$ in the curved valley, the time transformation gives $dt\sim \gamma_F dt^\prime$. 
When viewed in the lab frame, the time spent by the particle in the dissipation layer $\delta t$ is related to $\delta t^\prime\sim \gamma_F^2/\omega_B$ by 
\begin{equation}
\label{eq:delta_t1}
    \delta t\sim \gamma_F\delta t^\prime\sim\frac{ \gamma_F^3}{\omega_B},
\end{equation}
which is equivalent to $\delta t\sim \Delta/|\bar{v}_x|$ in the self-regulated layer with $Q\sim 1$.

The condition $Q\sim 1$ implies that the particle spends a large fraction of $\delta t^\prime$ with 
$v_z\sim -c$. The corresponding velocity
in the lab frame is found from $\gamma v_z=\gamma^\prime v_z^\prime\sim -\gamma^\prime c$, which gives 
\begin{equation}
\label{eq:vz_av}
    \bar{v}_z\sim - \frac{\gamma^\prime}{\bar{\gamma}}\,c\sim -\frac{c}{\bar{\gamma}^{1/2}}.
\end{equation}

\bibliography{ms}

\clearpage

\end{document}